\documentclass[a4paper,11pt]{article}

\usepackage{pos}
\usepackage{subfigure}

\newcommand{\A}{\mathrm{A}}
\newcommand{\B}{\mathrm{B}}
\newcommand{\p}{\partial}
\newcommand{\dg}{\dagger}
\newcommand{\ev}[1]{\langle  #1 \rangle }
\newcommand{\der}[2]{\frac{\mathrm{d} #1 }{\mathrm{d} #2 }}
\newcommand{\diff}{\mathrm{d}}

\DeclareMathOperator{\Tr}{Tr}

\title{Transverse momentum broadening in real-time lattice simulations of the glasma}
\author{A.~Ipp}
\author{D.~I.~Müller}
\author*{D.~Schuh}
\affiliation[]{Institute for Theoretical Physics, TU Wien, \\
	Wiedner Hauptstr. 8-10, 1040 Vienna, Austria}
\emailAdd{ipp@hep.itp.tuwien.ac.at}
\emailAdd{dmueller@hep.itp.tuwien.ac.at}
\emailAdd{schuh@hep.itp.tuwien.ac.at}

\abstract{The study of jets in heavy-ion collisions provides important information about the interaction of partons with the medium that they traverse. The seeds of jets are highly energetic partons, which are produced from hard scatterings during the collision event. As such, they are affected by all different stages of the medium's time evolution, including the glasma, which is the pre-equilibrium precursor state of the quark-gluon plasma. Here we  report on our numerical lattice simulations of partons traversing the boost-invariant, non-perturbative glasma as created at the early stages of collisions at RHIC and LHC. We find that partons quickly accumulate transverse momentum up to the saturation momentum during the glasma stage. Furthermore, we observe an interesting anisotropy in transverse momentum broadening of partons with larger broadening in the rapidity than in the azimuthal direction. Its origin can be related to correlations among the longitudinal color-electric and color-magnetic flux tubes in the initial state of the glasma. We compare these observations to the semi-analytic results obtained by a weak-field approximation, where we also find such an anisotropy in a parton's transverse momentum broadening.}

\FullConference{%
 The 38th International Symposium on Lattice Field Theory, LATTICE2021
  26th-30th July, 2021
  Zoom/Gather@Massachusetts Institute of Technology
}

\begin{document}
\maketitle

\section{Introduction}
The theoretical description of heavy-ion collisions poses many challenges, as it requires different schemes for the various stages of matter created in such a collision. An intensely researched stage is the quark-gluon plasma (QGP), which is well described by viscous hydrodynamics~\cite{Gale:2013, Romatschke:2017}. Because of the existence of a hydrodynamic attractor~\cite{Berges:2013}, bulk properties, such as particle multiplicities and flow harmonics~\cite{Schenke:2010, Schenke:2011, Schenke:2012, Gale:2012, Niemi:2015}, can be described without an exact knowledge of the initial conditions of the hydrodynamical evolution. However, probes, such as jets~\cite{Mehtar-Tani:2013, Connors:2017, Busza:2018}, which are created by hard scatterings during the collision, are affected by the evolution prior to the QGP. This pre-equilibrium precursor state, the glasma, is based on the color-glass condensate (CGC)~\cite{Gelis:2010, Gelis:2012}, which is an effective theory of high-energy QCD. In the glasma stage, expanding chromo-electric and chromo-magnetic flux tubes interact strongly with partons that lead to jets in later stages. Since the color fields are initially longitudinal~\cite{Lappi:2006}, this interaction is highly anisotropic. Jet quenching~\cite{Qin:2015} in heavy-ion collisions is sensitive to the very early stages~\cite{Andres:2019}. There, the momentum of said parton is influenced anisotropically by the color fields, which leads to an anisotropy in the momentum broadening of the subsequent jet.

Analytic calculations of glasma observables, such as momentum broadening, have proven to be difficult and, thus, necessitate some approximations. Oftentimes, the boost-invariant approximation is employed, which makes the system effectively 2+1 dimensional because the rapidity dependence is neglected. An intriguing limit for analytic considerations is the dilute limit~\cite{Kovner:1995}, in which the flux-tube color fields are deemed small. There, the momentum broadening of a test particle moving at constant speed can be calculated mostly analytically; the only numerics comes in when evaluating some integrals. Also, an interesting analytic relation between the anisotropy of the momentum broadening of a test parton and the correlations among the color flux tubes in the initial state of the glasma can be found. This was shown in~\cite{Ipp:2020a, Ipp:2020b}, which are the papers that these conference proceedings are based on. Recent advancements in the dilute limit have also been made in 3+1 dimensions~\cite{Ipp:2021}. Another limit, the lattice approximation, allows for strong color fields in the glasma, which is the more realistic case. In this real-time lattice field theoretic approach, the transverse plane is approximated by a square lattice with periodic boundary conditions, and the equations of motion, the (discretized) Yang-Mills equations, are solved with finite differences.

This paper is structured as follows: in sec.~\ref{sec:glasma}, we will discuss the theoretical framework of the glasma. Then, in sec.~\ref{sec:mom_broad}, we will introduce momentum broadening of a parton traveling through the glasma, which will be approximated in two ways: by a weak-field approximation (subsec.~\ref{subsec:weak-field_approx}) and by a lattice approximation (subsec.~\ref{subsec:lattice_approx}). We will present our results in sec.~\ref{sec:results_and_discussion} and conclude in sec.~\ref{sec:concl_and_outl}.

\section{The glasma} \label{sec:glasma}
The theoretical framework for the description of the glasma is the CGC. In the center-of-mass frame, partons of the two incoming nuclei A and B that carry high momentum are described by static, classical color charges. They move along the beam axis~$z$ and are described by the color current
\begin{equation}
    J^\mu_{(\A,\B)} = \delta^\mu_\pm \rho_{(\A,\B)}(x^\mp, \mathbf x),
\end{equation}
with the light-cone coordinates $x^\pm = (t \pm z) / \sqrt{2}$, $\mathbf x = (x, y)$ and the Kronecker delta~$\delta^\mu_\pm$. Their momentum is high enough so that they can be treated as infinitely thin~$\rho_{(\A,\B)}(x^\mp, \mathbf x) = \delta(x^\mp)\rho_{(\A,\B)}(\mathbf x)$, which makes this setup boost invariant, and their spatial extent in the transverse directions is large enough, so that we approximate them to be infinitely big in these directions. We use the McLerran-Venugopalan model~\cite{McLerran:1994a, McLerran:1994b} to describe the nuclei. In this model, the color charge density~$\rho$ is taken to be a random field that is distributed according to the Gaussian probability functional~$W[\rho]$, which is characterized by the charge density correlator
\begin{equation}
    \ev{\rho^a_{(\A, \B )}(x) \rho^b_{( \A, \B )}(y)} = (g \mu)^{2} \delta^{ab} \delta(x^{\mp} - y^{\mp}) \delta(x^{\mp}) \delta^{(2)}  (\mathbf x - \mathbf y), \label{eq:MV_charge_density_corr}
\end{equation}
where~$\mu$ is a model parameter that fixes the saturation momentum~$Q_s \propto g^2 \mu$. Hence, expectation values of observables are computed by taking the average over all configurations, weighted by~$W[\rho]$.

The high-momentum partons act as sources for the low-momentum partons. The latter are described by the classical vector field~$A^\mu$, and their relation is governed by the Yang-Mills equations
\begin{equation}
    D_\mu F^{\mu\nu}_{(\A,\B)} = J^\nu_{(\A,\B)}.
\end{equation}
In light-cone gauges ($A^+ = 0$ for nucleus~A and $A^- = 0$ for nucleus~B), the color field is purely transverse
\begin{equation}
    A^i_{(\A,\B)}(x^\mp, \mathbf x) = \frac{1}{ig} V_{(\A,\B)}(\mathbf x) \p^i V_{(\A,\B)}(\mathbf x) \theta(x^\mp), \label{eq:transverse_color_field}
\end{equation}
and the lightlike Wilson lines along the light-cone coordinate axes~$x^\pm$ are given by
\begin{equation}
    V^\dg_{(\A,\B)}(\mathbf x) = \mathcal{P} \exp{\left(i g \intop^{\infty}_{-\infty} dx^\mp \frac{\rho_{(\A,\B)} (x^\mp , \mathbf x)}{\mathbf \nabla^2 - m^2}\right)}, \label{eq:lightlike_Wilson_lines}
\end{equation}
with the infrared regulator~$m$ and the Yang-Mills coupling constant~$g$; $\mathcal{P}$ denotes path ordering. In Milne coordinates ($\tau = \sqrt{2 x^- x^+}$, $\eta = \ln (2 x^+ x^-)/2$) and temporal gauge ($A^\tau = 0$), the glasma initial conditions~\cite{Kovner:1995} read
\begin{align}
    A^i(\mathbf x) &= A^i_\A(\mathbf x) + A^i_\B(\mathbf x), \label{eq:g_ic1} \\
    A^\eta(\mathbf x) &= \frac{ig}{2} \left[ A^i_\A(\mathbf x), A^i_\B(\mathbf x)\right]. \label{eq:g_ic2}
\end{align}
Note that in the boost-invariant approximation, the Yang-Mills equations, which determine the evolution of the glasma, are source free.

\section{Momentum broadening in the glasma} \label{sec:mom_broad}
\begin{figure}
    \centering
    \includegraphics[scale=0.2]{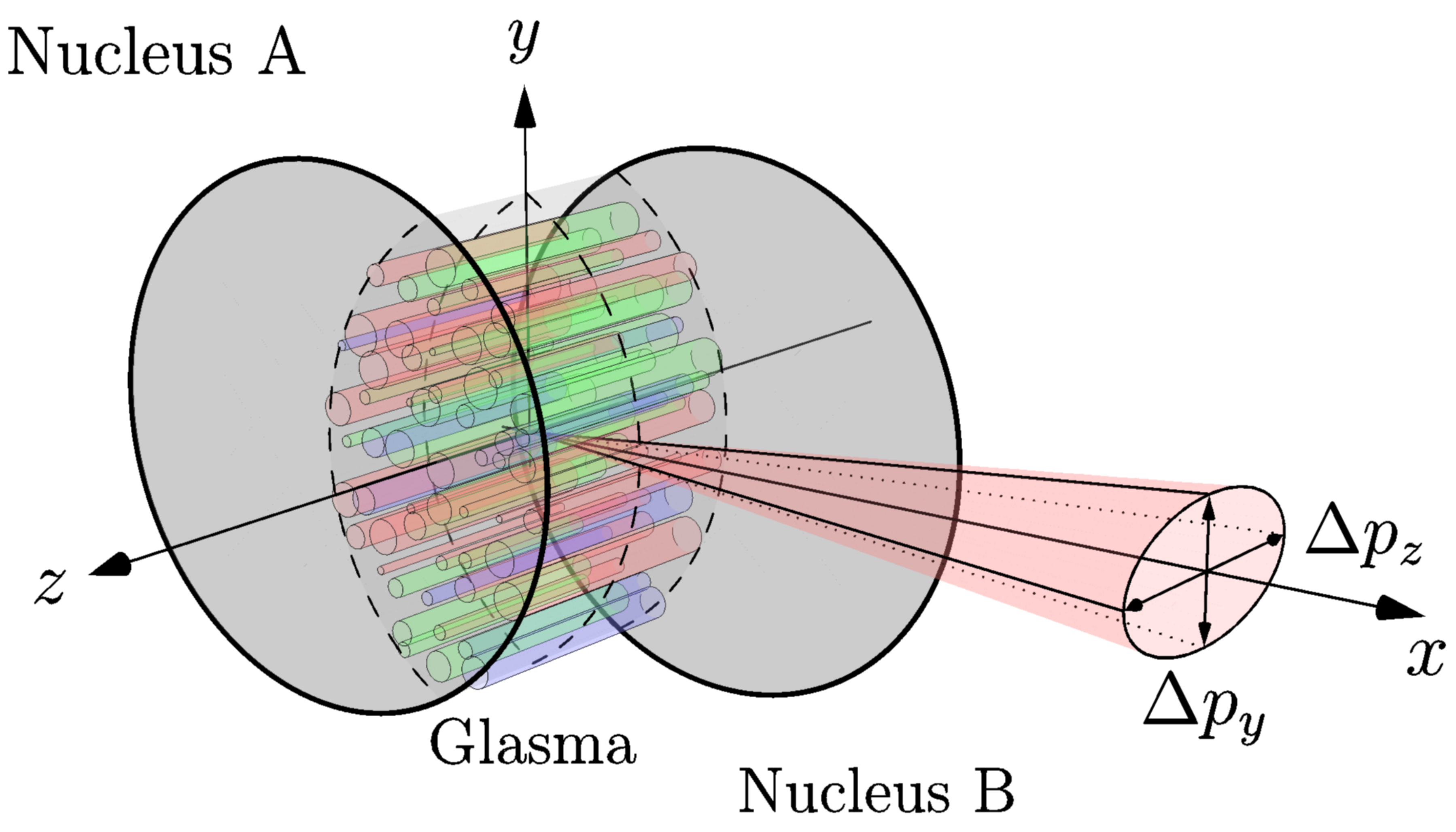}
    \caption{Schematic picture of a heavy-ion collision (taken from~\cite{Ipp:2020a}). When the two nuclei~A and~B, visualized in gray, collide, they produce the glasma, which is depicted by colorful (flux) tubes. A parton, the seed of a jet, is produced through hard scatterings during the same collision event. It moves in the $x$-direction, which is orthogonal to the beam axis ($z$-axis). Transverse momentum broadening refers to the $y$- and $z$-directions.}
    \label{fig:momentum_broadening}
\end{figure}

The momentum of a parton moving through the glasma is broadened because the color fields in the glasma exert strong Lorentz forces on the parton. We consider an ultra-relativistic parton that is created via hard scattering during the collision, which coincides with the origin of the coordinate system. The origin can be chosen without loss of generality because the transverse extent of the color sheets that approximate the nuclei is taken to be infinite (see sec.~\ref{sec:glasma}). The parton travels at the speed of light, and it is assumed to be too energetic to be deflected, but the color forces lead to an accumulation of momentum transverse to its trajectory, which we take to be the $x$-direction. Since the system is boost invariant, this can be done without loss of generality. Back-reactions from the parton to the glasma are neglected, i.e.~we consider the parton to be a test particle. Note that transverse momentum broadening refers to~$\ev{p^2_i}$ orthogonal to the particle trajectory, not the beam direction, i.e.~$i \in \{y,z\}$. The collision event is depicted in fig.~\ref{fig:momentum_broadening}.

The equations of motion for a test parton in a non-Abelian background field are given by the Wong equations
\begin{align}
	\der{p_\mu}{\tau} &= g Q^a(\tau) \der{x^\nu}{\tau} F^{a}_{\mu\nu}(\tau), \\
    \der{Q^a}{\tau} &= g \der{x^\mu}{\tau} f^{abc} A^b_\mu(\tau) Q^c(\tau),
\end{align}
with the trajectory~$x^\mu$ and the color charge of the quark~$Q^a$. The fields are evaluated along the particle trajectory. The solution of the Wong equations for a quark (indicated by the subscript~$q$) reads
\begin{equation}
	\ev{p^2_i(\tau)}_q = \frac{g^2}{N_c} \intop^\tau_0 \! \diff \tau' \intop^\tau_0 \! \diff \tau'' \, \ev{\mathrm{Tr} \left[ f^i(\tau') f^i(\tau'') \right]}, \label{eq:p_suqared}
\end{equation}
where~$f^i$ are functions that represent the color rotated Lorentz force
\begin{align}
	f^y(\tau) &= U(\tau) \left( E_y(\tau) - B_z(\tau) \right) U^\dg(\tau), \label{eq:f_y} \\
	f^z(\tau) &= U(\tau) \left( E_z(\tau) + B_y(\tau) \right) U^\dg(\tau), \label{eq:f_z}
\end{align}
and $U$ is a lightlike Wilson line in the fundamental representation along the particle trajectory
\begin{equation}
	U(\tau, 0) = \mathcal{P} \exp{\bigg( - i g \intop_0^\tau \! \diff \tau' \, A_x (\tau') \bigg)}. \label{eq:Wilson_line}
\end{equation}
The momentum broadening for an ultra-relativistic test gluon traveling through the boost-invariant glasma is related to eq.~\eqref{eq:p_suqared} via Casimir scaling
\begin{equation}
	\ev{p^2_i}_g = \frac{C_A}{C_F} \ev{p^2_i}_q,
\end{equation}
with the Casimirs~$C_A$ and~$C_F$ in the adjoint and fundamental representation, respectively.

Evaluating eq.~\eqref{eq:p_suqared} is challenging for two reasons: first, one would need to solve the Yang-Mills equations, and, second, it is a highly nonlinear functional of~$A_\mu$, as can be seen from eq.~\eqref{eq:Wilson_line}. Therefore, some approximations need to be made in order to proceed. In the following, we will discuss two possibilities: a weak-field approximation and a lattice approximation.

\subsection{Weak-field approximation} \label{subsec:weak-field_approx}
In the weak-field approximation~\cite{Kovner:1995}, we consider the charge density~$\rho$ that describes the nuclei to be small. As a consequence, the glasma initial conditions, which are given by eqs.~\eqref{eq:g_ic1} and~\eqref{eq:g_ic2}, can be treated perturbatively in~$\rho$, and the time evolution becomes Abelian in the lowest order that is not pure gauge. At this (fourth) order, the momentum broadening reads
\begin{equation}
    \ev{p^2_{(y, z)}(\tau)}_q = \int \! \frac{\diff^2 \mathbf k}{(2 \pi)^2} \, g(\tau, \mathbf k) \, c_{(E, B)}(k, m), \label{eq:p_squared_weak}
\end{equation}
where
\begin{equation}
    g(\tau, \mathbf k) = \frac{N_c^2 - 1}{2} g^8 \mu^4 \left| \intop^\tau_0 \! \diff \tau' \, \left( \frac{i k_x}{k} J_1(k \tau') + J_0(k \tau') \right) e^{i k_x \tau'} \right|^2
\end{equation}
is the same function for the $y$- and the $z$-component. It describes the time evolution of the glasma; $N_c$ pertains to the chosen gauge group SU$(N_c)$, $k$ is the norm of $\mathbf k$ and $J_n$ are the Bessel functions of the first kind. The components differ only by the initial time correlators~$c_B$ and~$c_E$
\begin{align}
    c_B(k, m) &= \int \! \frac{\diff^2 \mathbf p}{(2 \pi)^2} \, \frac{(\mathbf p \times \mathbf k)^2}{(p^2 + m^2)^2 (|\mathbf k - \mathbf p|^2 + m^2)^2}, \label{eq:cB_k} \\
    c_E(k, m) &= \int \! \frac{\diff^2 \mathbf p}{(2 \pi)^2} \, \frac{(\mathbf p \cdot (\mathbf k - \mathbf p))^2 }{(p^2 + m^2)^2 (|\mathbf k - \mathbf p|^2 + m^2)^2}. \label{eq:cE_k}
\end{align}

\subsection{Lattice approximation} \label{subsec:lattice_approx}
In the lattice approximation~\cite{Krasnitz:1998,Lappi:2003}, we discretize the transverse plane as a regular $N_T \times N_T$ lattice with periodic boundary conditions. The transverse lattice spacing is given by~$a_T$, and the transverse length of the lattice is~$L_T = N_T a_T$. The discrete proper times are given by~$\tau_n = n \Delta \tau$, with the time step~$\Delta \tau$ and $n \in \mathbb{N}$. The time step and the transverse lattice spacing are related by~$\Delta \tau = a_T / n_\tau$, where $n_\tau \ge 2$ is an even integer. The degrees of freedom are the gauge links in the transverse plane~$U_{x, \hat{i}}(\tau_n)$, the rapidity component of the gauge field~$A_{x, \eta}(\tau_n)$ and the conjugate momenta~$P^{i}_x(\tau_{n+1/2})$ and $P^{\eta}_x(\tau_{n+1/2})$. Note that the gauge links and the gauge field are evaluated at time steps~$\tau_n$, whereas the conjugate momenta are taken at fractional time steps~$\tau_{n+1/2}$. The Yang-Mills equations are replaced by a leapfrog scheme for finite time steps, where the covariant derivatives have been replaced by forward and backward differences. The glasma initial conditions, given by eqs.~\eqref{eq:g_ic1} and~\eqref{eq:g_ic2} also have to be discretized. Each individual nucleus is described by $N_s$~color sheets along the longitudinal direction, which accounts for path ordering~\cite{Fukushima:2007}. The charge density correlator, which is the discretized analog of eq.~\eqref{eq:MV_charge_density_corr}, then reads
\begin{equation} 
    \ev{\rho^a_{x,m} \rho^b_{y,n}} = \frac{ g^2 \mu^2 }{N_s a^2_T} \delta_{mn} \delta^{ab} \delta_{xy}.
\end{equation}
The indices~$m$ and $n$ refer to the respective color sheet. The numbers representing the discrete charge density~$\rho^a_{x,m}$ are drawn from a Gaussian distribution. They act as the inhomogeneity in the Poisson equations that have to be solved for each color sheet individually.\footnote{In the continuum, the solution of the Poisson equation in the transverse plane leads to eq.~\eqref{eq:transverse_color_field}.} The lightlike Wilson line~$V^\dg$ is approximated by a product over the individual color sheets, instead of given by a path ordered integral, as in eq.~\eqref{eq:lightlike_Wilson_lines}. They constitute the transverse gauge fields of the respective nucleus
\begin{equation}
    U^{(A,B)}_{x,\hat{i}} = V^{\vphantom{\dagger}}_{\vphantom{\hat{i}}(A,B),x} V^\dagger_{(A,B),x+\hat{i}},
\end{equation}
which are, in turn, the input for the glasma initial conditions.

In order to calculate momentum broadening, the functions~$f^i$ in eq.~\eqref{eq:p_suqared}, which are given by eqs.~\eqref{eq:f_y} and~\eqref{eq:f_z}, have to be discretized. We do this based on the ideas of the nearest-grid-point scheme of the colored-particle-in-cell method~\cite{Hu:1996, Moore:1997, Strickland:2007}: when the particle moves across the two-dimensional transverse lattice, its charge contributes to the lattice charge density only at the lattice site that is closest to it. Each time the nearest grid point changes, the color charge is color rotated with the Wilson line connecting the old nearest grid point with the new one, which is dictated by local gauge-covariant color-charge conservation.

In the setup of this paper, the particle moves in the $x$-direction, along the lightlike trajectory \mbox{$x^\mu = (\tau_n, \tau_n, 0, 0) + x_0^\mu$}, where~$x_0^\mu$ is its starting position at~$\tau_0$. At times~$t_n = n a_T = n_\tau \tau_n$, the particle position coincides with a lattice point, and at~$\bar{t}_n = (n + 1/2) a_T$, the nearest grid point of the particle changes, which is when its color charge color rotates. Therefore, the discretized versions of the Wilson lines~$U$ in eqs.~\eqref{eq:f_y} and~\eqref{eq:f_z}, which take care of the color rotation, are evaluated at times~$\bar{t}_n$. Their approximation reads
\begin{equation}
    U(0, \bar{t}_n) \approx U_{x_0, \hat{x}}(\bar{t}_0) \mspace{2mu} U_{x_1, \hat{x}}(\bar{t}_1) \; \dots \; U_{x_{n-1}, \hat{x}}(\bar{t}_{n-1}) \mspace{2mu} U_{x_n, \hat{x}}(\bar{t}_n).
\end{equation}
We approximate the (discretized) color-electric and color-magnetic fields of eqs.~\eqref{eq:f_y} and~\eqref{eq:f_z} to second order in both lattice spacing and time step. They are evaluated at times~$t_n$ at the lattice site at which the parton is located. Gauge links and color fields are, thus, evaluated at different times, and the lattice versions of eqs.~\eqref{eq:f_y} and~\eqref{eq:f_z} read
\begin{align}
    f^y(t_n) &= U(0, \bar{t}_n) \left( E_y(t_n) - B_z(t_n) \right) U(\bar{t}_n, 0), \\
    f^z(t_n) &= U(0, \bar{t}_n) \left( E_z(t_n) + B_y(t_n) \right) U(\bar{t}_n, 0).
\end{align}
The time integrals that are needed to calculate momentum broadening are approximated by sums, and, therefore, the lattice approximation of eq.~\eqref{eq:p_suqared} is given by
\begin{equation}
    \ev{p^2_{(y,z)}(t_n)}_q \approx \frac{g^2 a_T^2}{N_c} \ev{ \Tr \bigg[ \big(\sum^n_{i=0} f^{(y, z)}(t_n) \big)^2 \bigg] }. \label{eq:p_suqared_latt}
\end{equation}
The average is performed over the random color charge densities that are used in the initial conditions.

\section{Results and discussion} \label{sec:results_and_discussion}
In this section, we evaluate eq.~\eqref{eq:p_suqared_latt} with SU$(3)$ real-time lattice simulations for a dense glasma. The density is determined by the ratio~$m / g^2 \mu$: it is small for a dense glasma and large for a dilute glasma. We work at fixed saturation momentum~$Q_s$ by choosing $g^2 \mu$ for some fixed saturation momentum according to the numerical results presented in~\cite{Lappi:2007}. The transverse plane is approximated by~$N_T = 1024$ points in each direction, and the time step is related to the lattice spacing via~$\Delta \tau = a_T / 16$. Each nucleus is approximated by $N_s = 50$ color sheets. The lattice resolution $g^2 \mu a_T \approx 0.1$ for  $g^2\mu L = 100$ is sufficient to resolve glasma flux tubes, which have a diameter of $Q^{-1}_s \approx (g^2 \mu)^{-1}$~\cite{Lappi:2006}. We choose a few different values for the infrared regulator~$m$, so that the glasma that we simulate is dense: $m / g^2 \mu \in \{ 0, 0.05, 0.1, 0.2 \}$. For a vanishing infrared regulator, we implement color neutrality at the size of the system~$L$ by eliminating the zero mode of the charge density~\cite{Krasnitz:2003}. For a non-vanishing infrared regulator, the system is large enough to resolve multiple color neutral domains.

\begin{figure}
    \centering
    \subfigure[Momentum broadening at early times]{\includegraphics[height=4.85cm]{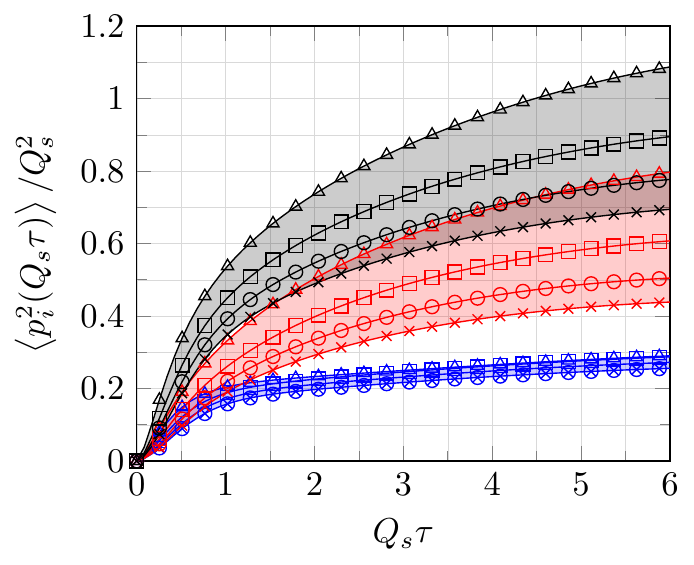}}
    \quad
    \subfigure[Momentum broadening up to later times]{\includegraphics[height=4.85cm]{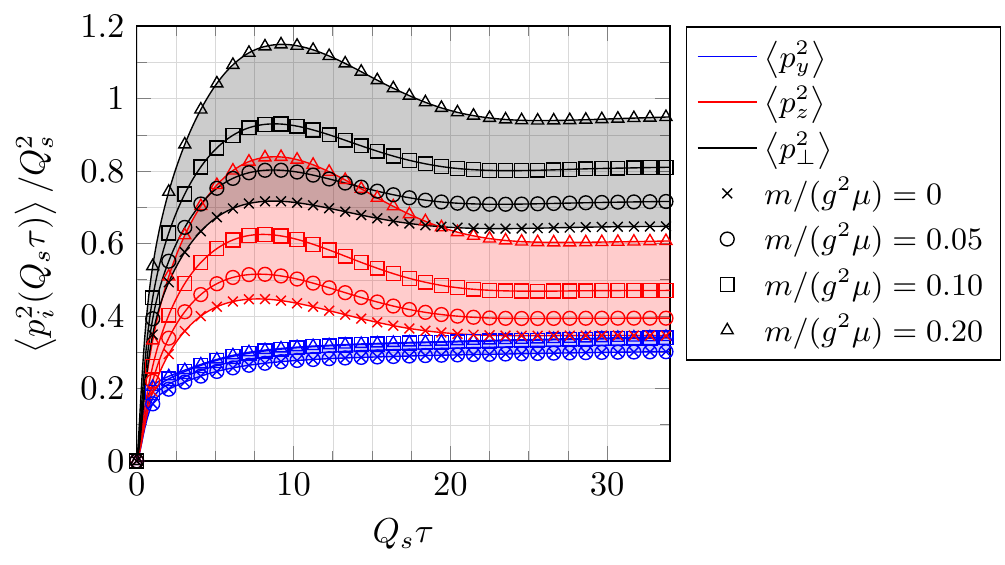}}
    \caption{Accumulated transverse momentum for a high-energy quark moving in the $x$-direction as a function of the dimensionless proper time $Q_s \tau$ from SU$(3)$ real-time lattice simulations of the dense glasma (taken from \cite{Ipp:2020a}). The symbols correspond to different values of $m / g^2 \mu$; the bands indicate the ranges that $\ev{p^2_y}$, $\ev{p^2_z}$ and $\ev{p^2_\perp} = \ev{p^2_y} + \ev{p^2_z}$ can take for the chosen values for this ratio. The results have been averaged over 50 random initial conditions. The $y$- and the $z$-component exhibit different behavior, which leads to an anisotropy \mbox{$\ev{p^2_z} / \ev{p^2_y} \neq 1$} in momentum broadening.}
    \label{fig:acc_mom_tau}
\end{figure}

The main results of these lattice simulations are presented in fig.~\ref{fig:acc_mom_tau}. It depicts transverse momentum broadening for quarks as a function of dimensionless proper time~$Q_s \tau$, more precisely, it shows the broadening within the plane transverse to the beam axis ($y$-component), the broadening along the beam axis ($z$-component) and the total transverse momentum broadening. The behavior of the $y$-component and the $z$-component are quite different: the former barely depends on the density of the glasma, rises sharply until~$\tau \approx Q_s^{-1}$ and flattens out afterwards, whereas the latter depends strongly on this density, exhibits a peak just below~$Q_s \tau = 10$ and decreases afterwards. Naturally, the total transverse momentum broadening inherits the dependence on the ratio~$m / g^2 \mu$ and the peak below~$Q_s \tau = 10$ from $\ev{p^2_z}$, since it is simply the sum of its individual components, and $\ev{p^2_y}$ becomes rather flat quite quickly. A momentum broadening anisotropy starts to show around~$\tau \approx Q_s^{-1}$ and keeps growing until~$\ev{p^2_z}$ reaches its peak. Then, it starts to shrink again and vanishes almost entirely for~$m / g^2 \mu = 0$. However, the shrinking of momentum broadening along the beam axis and the associated shrinking of the anisotropy occur at late times~$Q_s \tau \approx 10$, where the glasma is no longer a valid description of the state of matter that is produced in heavy-ion collisions.

\begin{figure}
    \centering
    \includegraphics[height=8.0cm]{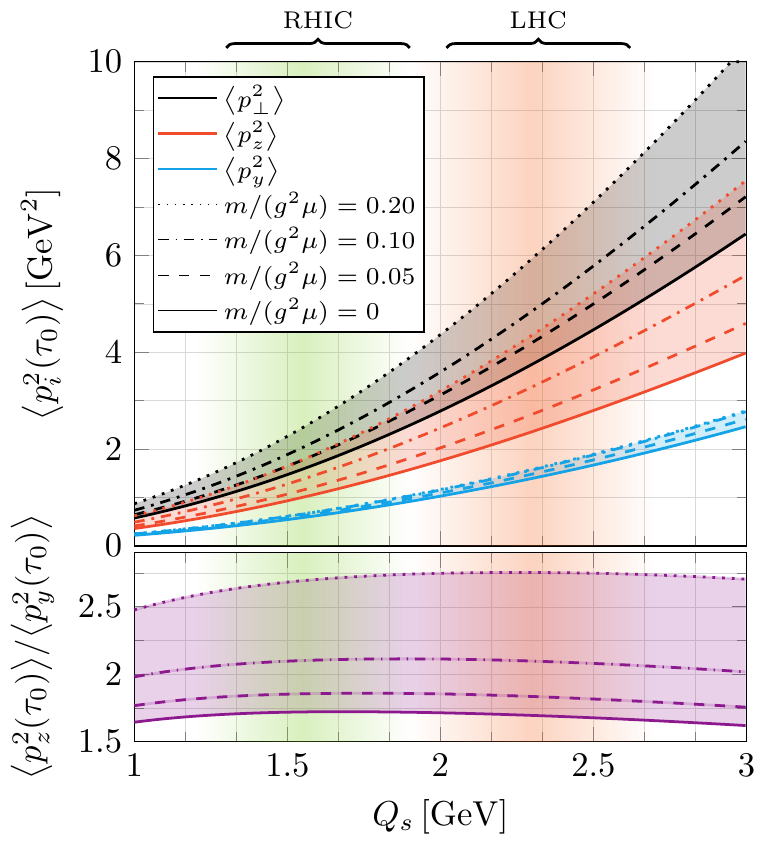}
    \caption{Accumulated transverse momentum (top) and momentum broadening anisotropy (bottom) for a high-energy quark moving in the $x$-direction as a function of the saturation momentum~$Q_s$ at $\tau_0 = 0.6 \, \mathrm{fm}/\mathrm{c}$ (taken from~\cite{Ipp:2020b}). The different line styles indicate the different values of $m / g^2 \mu$; the bands display the ranges in which the respective components lie. The vertical bands in green and red illustrate the relevant regions of the saturation momentum for RHIC and LHC, respectively.}
    \label{fig:acc_mom_Qs}
\end{figure}

A typical starting time of jet energy loss calculations that neglect pre-equilibrium effects~\cite{Andres:2019} is \mbox{$\tau_0 = 0.6 \, \mathrm{fm}/\mathrm{c}$}. Figure~\ref{fig:acc_mom_Qs} displays the accumulated momentum up to this point in time for different values of~$Q_s$ and the corresponding anisotropy in momentum broadening. The upper plot shows the total transverse momentum broadening \mbox{$\ev{p^2_\perp} \approx Q_s^2$}, although its exact value depends on the density of the glasma. It is apparent, as it was in fig.~\ref{fig:acc_mom_tau}, that this dependence comes largely from the $z$-component. The lower plot reveals that, while depending on the density of the glasma, the anisotropy is largely independent of the saturation momentum.

\begin{figure}
    \centering
    \includegraphics[height=4.85cm]{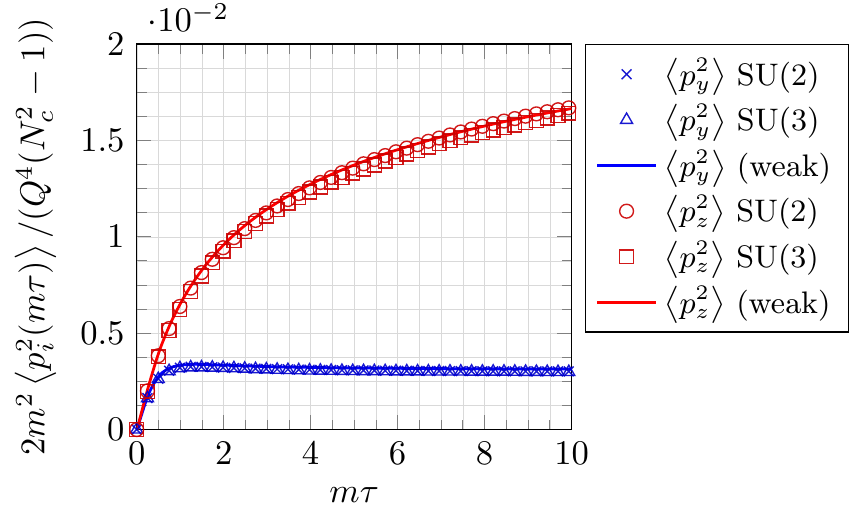}
    \caption{Accumulated transverse momentum in the dilute glasma: comparison of the weak-field approximation and the lattice approximation in the dilute limit (taken from~\cite{Ipp:2020a}). The lattice simulations (symbols) have been performed with SU$(2)$ and SU$(3)$ as gauge groups, the results scale with $N_c^2 - 1$ and they agree with the semi-analytic weak-field approximation. The behavior of both components is similar to the early-time behavior in the dense glasma, also leading to a considerable anisotropy.}
    \label{fig:acc_mom_weak}
\end{figure}

The next plot, fig.~\ref{fig:acc_mom_weak}, compares the lattice approximation of a dilute glasma $m / g^2 \mu \gg 1$ to the weak-field approximation, which amounts to the numerical evaluation of eq.~\eqref{eq:p_squared_weak}. This provides a consistency check for both the analytic approximations made during the derivation of~eq.~\eqref{eq:p_squared_weak} and the implementation of the lattice simulation with SU$(2)$ and SU$(3)$. Furthermore, it shows that the different behavior of the two components~$\ev{p^2_y}$ and~$\ev{p^2_z}$ and, therefore, the anisotropy in momentum broadening are also present in the dilute glasma, where we have an analytic relation, namely eq.~\eqref{eq:p_squared_weak}, between the momentum broadening and the initial time correlators, which are given by eqs.~\eqref{eq:cB_k} and~\eqref{eq:cE_k}. The difference between color magnetic and color electric flux tubes is, thus, responsible for the anisotropy in the dilute glasma. Notably, we do not have such a relation in the dense glasma, which is the physically relevant one, but momentum broadening in the dilute glasma, shown in fig.~\ref{fig:acc_mom_weak}, looks qualitatively similar to the early time behavior of momentum broadening in the dense glasma, depicted in fig.~\ref{fig:acc_mom_tau}. The unphysical dilute glasma may, thus, give us some intuition, nevertheless.

\section{Conclusions and outlook} \label{sec:concl_and_outl}
In this paper, we discussed momentum broadening of a high-energy test quark or gluon in the pre-equilibrium stage of the QGP, the glasma. After presenting two approximations that can be used to evaluate said broadening numerically, we analyzed its time evolution for fixed saturation momentum and its dependence on the saturation momentum at a fixed proper time. We chose this time to be $\tau_0 = 0.6 \, \mathrm{fm}/\mathrm{c}$. This corresponds to the starting time of jet energy loss calculations that neglect pre-equilibrium effects, and, thus, the presented momentum broadening amounts to the total accumulated momentum broadening before that.

Additionally, we compared the aforementioned two approximations in the dilute glasma, where they should and do coincide. In the dilute limit, the momentum broadening of a high energy test parton can be linked analytically to the initial stages of the glasma. This way, the anisotropy can be explained by the difference of the color-electric and color-magnetic flux tubes at the creation of the glasma. We point out the qualitative similarity between momentum broadening in the dilute glasma and the early-time behavior of the dense glasma.

The work presented in this paper can be extended in multiple ways: first, one could use more realistic initial conditions for the glasma. These would include the finite spatial extent of the nuclei in the transverse plane and would allow for the study of off-central collisions. Furthermore, the nuclei could be treated as finitely thick in the beam direction. This relaxation implies a rapidity dependence of the system. Progress in this direction has been made recently in~\cite{Ipp:2021}. Second, one could relax the ultra-relativistic-test-particle approximation. This relaxation would imply a deflection of the parton and its back-reaction on the glasma. Finally, one could study other observables in the glasma, e.g.~the energy loss, which is not accounted for in the approximations made in this work.

\acknowledgments
This work has been supported by the Austrian Science Fund FWF No.~P32446-N27 and No.~P28352. 
The Titan\,V GPU used for this research was donated by the NVIDIA Corporation.

\bibliographystyle{JHEP.bst}
\bibliography{references.bib}

\providecommand{\href}[2]{#2}\begingroup\raggedright\begin{thebibliography}{10}

\bibitem{Gale:2013}
C.~Gale, S.~Jeon and B.~Schenke, \emph{{Hydrodynamic Modeling of Heavy-Ion
  Collisions}}, \href{https://doi.org/10.1142/S0217751X13400113}{\emph{Int. J.
  Mod. Phys.} {\bfseries A28} (2013) 1340011}
  [\href{https://arxiv.org/abs/1301.5893}{{\ttfamily 1301.5893}}].

\bibitem{Romatschke:2017}
P.~Romatschke and U.~Romatschke, \emph{{Relativistic Fluid Dynamics In and Out
  of Equilibrium}}, Cambridge Monographs on Mathematical Physics, Cambridge
  University Press (2019),
  \href{https://doi.org/10.1017/9781108651998}{10.1017/9781108651998},
  [\href{https://arxiv.org/abs/1712.05815}{{\ttfamily 1712.05815}}].

\bibitem{Berges:2013}
J.~Berges, K.~Boguslavski, S.~Schlichting and R.~Venugopalan, \emph{{Universal
  attractor in a highly occupied non-Abelian plasma}},
  \href{https://doi.org/10.1103/PhysRevD.89.114007}{\emph{Phys. Rev.}
  {\bfseries D89} (2014) 114007}
  [\href{https://arxiv.org/abs/1311.3005}{{\ttfamily 1311.3005}}].

\bibitem{Schenke:2010}
B.~Schenke, S.~Jeon and C.~Gale, \emph{{(3+1)D hydrodynamic simulation of
  relativistic heavy-ion collisions}},
  \href{https://doi.org/10.1103/PhysRevC.82.014903}{\emph{Phys. Rev.}
  {\bfseries C82} (2010) 014903}
  [\href{https://arxiv.org/abs/1004.1408}{{\ttfamily 1004.1408}}].

\bibitem{Schenke:2011}
B.~Schenke, S.~Jeon and C.~Gale, \emph{{Higher flow harmonics from (3+1)D
  event-by-event viscous hydrodynamics}},
  \href{https://doi.org/10.1103/PhysRevC.85.024901}{\emph{Phys. Rev.}
  {\bfseries C85} (2012) 024901}
  [\href{https://arxiv.org/abs/1109.6289}{{\ttfamily 1109.6289}}].

\bibitem{Schenke:2012}
B.~Schenke, P.~Tribedy and R.~Venugopalan, \emph{{Event-by-event gluon
  multiplicity, energy density, and eccentricities in ultrarelativistic
  heavy-ion collisions}},
  \href{https://doi.org/10.1103/PhysRevC.86.034908}{\emph{Phys. Rev.}
  {\bfseries C86} (2012) 034908}
  [\href{https://arxiv.org/abs/1206.6805}{{\ttfamily 1206.6805}}].

\bibitem{Gale:2012}
C.~Gale, S.~Jeon, B.~Schenke, P.~Tribedy and R.~Venugopalan,
  \emph{{Event-by-event anisotropic flow in heavy-ion collisions from combined
  Yang-Mills and viscous fluid dynamics}},
  \href{https://doi.org/10.1103/PhysRevLett.110.012302}{\emph{Phys. Rev. Lett.}
  {\bfseries 110} (2013) 012302}
  [\href{https://arxiv.org/abs/1209.6330}{{\ttfamily 1209.6330}}].

\bibitem{Niemi:2015}
H.~Niemi, K.J.~Eskola and R.~Paatelainen, \emph{{Event-by-event fluctuations in
  a perturbative QCD + saturation + hydrodynamics model: Determining QCD matter
  shear viscosity in ultrarelativistic heavy-ion collisions}},
  \href{https://doi.org/10.1103/PhysRevC.93.024907}{\emph{Phys. Rev.}
  {\bfseries C93} (2016) 024907}
  [\href{https://arxiv.org/abs/1505.02677}{{\ttfamily 1505.02677}}].

\bibitem{Mehtar-Tani:2013}
Y.~Mehtar-Tani, J.G.~Milhano and K.~Tywoniuk, \emph{{Jet physics in heavy-ion
  collisions}}, \href{https://doi.org/10.1142/S0217751X13400137}{\emph{Int. J.
  Mod. Phys.} {\bfseries A28} (2013) 1340013}
  [\href{https://arxiv.org/abs/1302.2579}{{\ttfamily 1302.2579}}].

\bibitem{Connors:2017}
M.~Connors, C.~Nattrass, R.~Reed and S.~Salur, \emph{{Jet measurements in heavy
  ion physics}}, \href{https://doi.org/10.1103/RevModPhys.90.025005}{\emph{Rev.
  Mod. Phys.} {\bfseries 90} (2018) 025005}
  [\href{https://arxiv.org/abs/1705.01974}{{\ttfamily 1705.01974}}].

\bibitem{Busza:2018}
W.~Busza, K.~Rajagopal and W.~van~der Schee, \emph{{Heavy Ion Collisions: The
  Big Picture, and the Big Questions}},
  \href{https://doi.org/10.1146/annurev-nucl-101917-020852}{\emph{Ann. Rev.
  Nucl. Part. Sci.} {\bfseries 68} (2018) 339}
  [\href{https://arxiv.org/abs/1802.04801}{{\ttfamily 1802.04801}}].

\bibitem{Gelis:2010}
F.~Gelis, E.~Iancu, J.~Jalilian-Marian and R.~Venugopalan, \emph{{The Color
  Glass Condensate}},
  \href{https://doi.org/10.1146/annurev.nucl.010909.083629}{\emph{Ann. Rev.
  Nucl. Part. Sci.} {\bfseries 60} (2010) 463}
  [\href{https://arxiv.org/abs/1002.0333}{{\ttfamily 1002.0333}}].

\bibitem{Gelis:2012}
F.~Gelis, \emph{{Color Glass Condensate and Glasma}},
  \href{https://doi.org/10.1142/S0217751X13300019}{\emph{Int. J. Mod. Phys.}
  {\bfseries A28} (2013) 1330001}
  [\href{https://arxiv.org/abs/1211.3327}{{\ttfamily 1211.3327}}].

\bibitem{Lappi:2006}
T.~Lappi and L.~McLerran, \emph{{Some features of the glasma}},
  \href{https://doi.org/10.1016/j.nuclphysa.2006.04.001}{\emph{Nucl. Phys.}
  {\bfseries A772} (2006) 200}
  [\href{https://arxiv.org/abs/hep-ph/0602189}{{\ttfamily hep-ph/0602189}}].

\bibitem{Qin:2015}
G.-Y.~Qin and X.-N.~Wang, \emph{{Jet quenching in high-energy heavy-ion
  collisions}}, \href{https://doi.org/10.1142/S0218301315300143,
  10.1142/9789814663717_0007}{\emph{Int. J. Mod. Phys.} {\bfseries E24} (2015)
  1530014} [\href{https://arxiv.org/abs/1511.00790}{{\ttfamily 1511.00790}}].

\bibitem{Andres:2019}
C.~Andres, N.~Armesto, H.~Niemi, R.~Paatelainen and C.A.~Salgado, \emph{{Jet
  quenching as a probe of the initial stages in heavy-ion collisions}},
  \href{https://doi.org/10.1016/j.physletb.2020.135318}{\emph{Phys. Lett. B}
  {\bfseries 803} (2020) 135318}
  [\href{https://arxiv.org/abs/1902.03231}{{\ttfamily 1902.03231}}].

\bibitem{Kovner:1995}
A.~Kovner, L.D.~McLerran and H.~Weigert, \emph{{Gluon production at high
  transverse momentum in the McLerran-Venugopalan model of nuclear structure
  functions}}, \href{https://doi.org/10.1103/PhysRevD.52.3809}{\emph{Phys.
  Rev.} {\bfseries D52} (1995) 3809}
  [\href{https://arxiv.org/abs/hep-ph/9505320}{{\ttfamily hep-ph/9505320}}].

\bibitem{Ipp:2020a}
A.~Ipp, D.~M\"uller and D.~Schuh, \emph{Anisotropic momentum broadening in the
  2+1d glasma: Analytic weak field approximation and lattice simulations},
  \href{https://doi.org/10.1103/physrevd.102.074001}{\emph{Physical Review D}
  {\bfseries 102} (2020) } [\href{https://arxiv.org/abs/2001.10001}{{\ttfamily
  2001.10001}}].

\bibitem{Ipp:2020b}
A.~Ipp, D.~M\"uller and D.~Schuh, \emph{Jet momentum broadening in the
  pre-equilibrium glasma},
  \href{https://doi.org/10.1016/j.physletb.2020.135810}{\emph{Physics Letters
  B} {\bfseries 810} (2020) 135810}
  [\href{https://arxiv.org/abs/2009.14206}{{\ttfamily 2009.14206}}].

\bibitem{Ipp:2021}
A.~Ipp, D.I.~Müller, S.~Schlichting and P.~Singh, \emph{Space-time structure
  of 3+1d color fields in high energy nuclear collisions},
  \href{https://arxiv.org/abs/2109.05028}{{\ttfamily 2109.05028}}.

\bibitem{McLerran:1994a}
L.D.~McLerran and R.~Venugopalan, \emph{{Gluon distribution functions for very
  large nuclei at small transverse momentum}},
  \href{https://doi.org/10.1103/PhysRevD.49.3352}{\emph{Phys. Rev.} {\bfseries
  D49} (1994) 3352} [\href{https://arxiv.org/abs/hep-ph/9311205}{{\ttfamily
  hep-ph/9311205}}].

\bibitem{McLerran:1994b}
L.~{McLerran} and R.~{Venugopalan}, \emph{{Computing quark and gluon
  distribution functions for very large nuclei}},
  \href{https://doi.org/10.1103/PhysRevD.49.2233}{\emph{PRD} {\bfseries 49}
  (1994) 2233} [\href{https://arxiv.org/abs/hep-ph/9309289}{{\ttfamily
  hep-ph/9309289}}].

\bibitem{Krasnitz:1998}
A.~Krasnitz and R.~Venugopalan, \emph{{Nonperturbative computation of gluon
  minijet production in nuclear collisions at very high-energies}},
  \href{https://doi.org/10.1016/S0550-3213(99)00366-1}{\emph{Nucl. Phys.}
  {\bfseries B557} (1999) 237}
  [\href{https://arxiv.org/abs/hep-ph/9809433}{{\ttfamily hep-ph/9809433}}].

\bibitem{Lappi:2003}
T.~Lappi, \emph{{Production of gluons in the classical field model for heavy
  ion collisions}},
  \href{https://doi.org/10.1103/PhysRevC.67.054903}{\emph{Phys.Rev.} {\bfseries
  C67} (2003) 054903} [\href{https://arxiv.org/abs/hep-ph/0303076}{{\ttfamily
  hep-ph/0303076}}].

\bibitem{Fukushima:2007}
K.~Fukushima, \emph{{Randomness in infinitesimal extent in the
  McLerran-Venugopalan model}},
  \href{https://doi.org/10.1103/PhysRevD.77.074005}{\emph{Phys. Rev.}
  {\bfseries D77} (2008) 074005}
  [\href{https://arxiv.org/abs/0711.2364}{{\ttfamily 0711.2364}}].

\bibitem{Hu:1996}
C.~Hu and B.~M\"uller, \emph{{Classical lattice gauge field with hard thermal
  loops}}, \href{https://doi.org/10.1016/S0370-2693(97)00851-4}{\emph{Phys.
  Lett.} {\bfseries B409} (1997) 377}
  [\href{https://arxiv.org/abs/hep-ph/9611292}{{\ttfamily hep-ph/9611292}}].

\bibitem{Moore:1997}
G.D.~Moore, C.~Hu and B.~M\"uller, \emph{{Chern-Simons number diffusion with
  hard thermal loops}},
  \href{https://doi.org/10.1103/PhysRevD.58.045001}{\emph{Phys. Rev.}
  {\bfseries D58} (1998) 045001}
  [\href{https://arxiv.org/abs/hep-ph/9710436}{{\ttfamily hep-ph/9710436}}].

\bibitem{Strickland:2007}
A.~Dumitru, Y.~Nara and M.~Strickland, \emph{{Ultraviolet avalanche in
  anisotropic non-Abelian plasmas}},
  \href{https://doi.org/10.1103/PhysRevD.75.025016}{\emph{Phys. Rev.}
  {\bfseries D75} (2007) 025016}
  [\href{https://arxiv.org/abs/hep-ph/0604149}{{\ttfamily hep-ph/0604149}}].

\bibitem{Lappi:2007}
T.~Lappi, \emph{{Wilson line correlator in the MV model: Relating the glasma to
  deep inelastic scattering}},
  \href{https://doi.org/10.1140/epjc/s10052-008-0588-4}{\emph{Eur. Phys. J.}
  {\bfseries C55} (2008) 285}
  [\href{https://arxiv.org/abs/0711.3039}{{\ttfamily 0711.3039}}].

\bibitem{Krasnitz:2003}
A.~Krasnitz, Y.~Nara and R.~Venugopalan, \emph{{Gluon production in the color
  glass condensate model of collisions of ultrarelativistic finite nuclei}},
  \href{https://doi.org/10.1016/S0375-9474(03)00636-5}{\emph{Nucl. Phys.}
  {\bfseries A717} (2003) 268}
  [\href{https://arxiv.org/abs/hep-ph/0209269}{{\ttfamily hep-ph/0209269}}].

\end{thebibliography}\endgroup

\end{document}